# SOI technology for quantum information processing

S. De Franceschi[1,2,3], L. Hutin[1,2,3], R. Maurand[1,2,3], L. Bourdet[1,2,3], H. Bohuslavskyi[1,2,3], A. Corna[1,2,3],
D. Kotekar-Patil[1,2,3], S. Barraud[1,2,3], X. Jehl[1,2,3], Y.-M. Niquet[1,2,3], M. Sanquer[1,2,3], and M. Vinet[1,2,3]
[1]CEA, INAC, F-38000 Grenoble, France, email: silvano.defranceschi@cea.fr
[2]CEA, LETI, Minatec Campus, F-38054 Grenoble, France
[3]University Grenoble Alpes, F-38000 Grenoble, France

*Abstract*— We present recent progress towards the implementation of a scalable quantum processor based on fully-depleted silicon-on-insulator (FDSOI) technology. In particular, we discuss an approach where the elementary bits of quantum information – so-called qubits – are encoded in the spin degree of freedom of gate-confined holes in p-type devices. We show how a hole-spin can be efficiently manipulated by means of a microwave excitation applied to the corresponding confining gate. The hole spin state can be read out and reinitialized through a Pauli blockade mechanism. The studied devices are derived from silicon nanowire field-effect transistors. We discuss their prospects for scalability and, more broadly, the potential advantages of FDSOI technology.

## I. Introduction

Quantum computation requires all qubits of a quantum processor to be individually addressed, i.e. initialized, manipulated (through either single-qubit or two-qubit operations), and measured [1]. When it comes to up-scaling to larger numbers of integrated qubits (e.g. millions of qubits are envisioned in the case of fault-tolerant surface-code architectures [2]), this requirement becomes utterly demanding. Among all solid-state implementations, silicon-based qubits are particularly attractive in view of the possibility to leverage the well-established engineering and integration capabilities of microelectronics technology. At the same time, it has been recently shown that silicon spin qubits can exhibit outstanding performances in terms of quantum coherence and fidelity, especially when natural silicon is replaced by nuclear-spin-free $^{28}$Si [3]. These groundbreaking achievements on one side, and the latest advances in CMOS device scaling and pitch miniaturization on the other, make the development of spin qubits on an industry-standard CMOS platform a timely and viable opportunity. We present here an approach relying on 300-mm FDSOI technology.

## II. First Hole-Spin Qubit

All silicon-based qubits reported so far rely on either electron or nuclear spins. Single-qubit operations are performed by means of a local magnetic-field modulation. This is obtained either by an oscillating current through a locally defined metal strip [3] or by moving the electron qubit in the stray magnetic field of a nearby micro-magnet [4]. In the case of holes, their inherently strong spin-orbit coupling offers the opportunity to perform single-qubit operations simply by a time-controlled microwave modulation of a gate voltage [5], thereby eliminating the need for additional qubit-control elements.

In order to demonstrate hole-spin qubit functionality we used a double quantum dot (QD) device based on SOI nanowire MOSFET technology [6] (Fig. 1). The device (Fig. 2) has an undoped silicon channel where two QDs, QD1 and QD2, are formed by hole accumulation under the two gates, gate 1 and gate 2, respectively. Transport across the double QD is primarily dominated by the Coulomb blockade (CB) effect [7]. As a result, holes tunnel one-by-one across the two QDs only for those combinations of gate voltages, $V_{G1}$ and $V_{G2}$, for which the electrochemical potentials of the two QDs are aligned with each other and fall within the source-drain bias window. In addition, a spin blockade (SB) effect occurs when inter-dot tunneling requires a triplet-to-singlet transition [7], as in the case of the (1,1) → (0,2) transition, where the first and second digits represent the number of holes on QD1 and QD2, respectively. In this SB regime, qubit functionality can be established as illustrated in Fig. 3. We apply to Gate 1 a voltage modulation consisting of a rectangular pulse and a superimposed microwave (MW) burst whose frequency is tuned to resonance with the spin splitting in QD1, which is induced by a static magnetic field of the order of 0.1 T. Starting from the SB condition, where the spin state in QD1 (i.e. the hole qubit) is initialized (e.g. to state |↑⟩), a negative pulse brings the device from SB to CB where tunneling is forbidden by a Coulomb energy barrier regardless of the spin orientation. In this regime, a microwave burst of duration τ induces a coherent Rabi oscillation of the QD1 spin, turning its state into a superposition cos(θ/2)|↑⟩+ sin(θ/2)|↓⟩, with the rotation angle θ = 2π $f_R$ τ, where $f_R$ is the so-called Rabi frequency. When $V_{G1}$ is raised back to the original value, the rotated hole spin will be allowed to tunnel to QD2 with a probability proportional to its |↓⟩ component, i.e. to $\sin^2(θ/2)$. This translates into measurable current oscillations as shown in Fig. 4. The oscillation frequency $f_R$ increases linearly with the MW voltage excitation. The oscillation amplitude decays with τ, reflecting spin dephasing. From Ramsey-fringes interference and Hann-echo experiments [8] (not shown) we estimate an inhomogeneous dephasing time $T_2^* = 60$ ns, and an intrinsic decoherence time $T_2 = 250$ ns. While easy to implement and useful for proof-of-concept qubit demonstrations, the transport-based spin readout technique discussed here above presents significant limitations: 1) it does not allow single-shot readout, i.e. the ability to perform individual spin measurements; 2) it is not scalable. In the next section, we discuss an alternative two-gate geometry for which both of these limitations could be overcome.

## III. CORNER QUANTUM DOTS IN SPLIT-GATE DEVICES

In the narrow-channel device of Fig. 2, the holes are localized under the Gates close to the top face of the Si nanowire (see Fig. 5a). For larger channel widths, however, two pronounced potential minima develop at the upper nanowire corners leading to a pair of clearly distinct QDs [9] (see Fig. 5b).

Splitting the single Gate into two face-to-face Gates (see Fig. 6a,b) enables independent electrostatic control of the two corner dots. The inter-dot coupling, mediated by tunneling and Coulomb interaction, can be further tuned by means of a back-gate voltage applied to the handle silicon layer. In Figs. 7 and 8, we provide experimental evidence of this tunability. The corresponding data sets refer to low-temperature transport measurements in an n-type split-gate device with W = 45 nm. The measured current is due to single-electron tunneling through the two corner dots in parallel. For $V_{BG}$ = -15 V (Fig. 7), the two dots are only weakly coupled. Current ridges with almost vertical (horizontal) slopes denote the addition of electrons to QD1 (QD2). For $V_{BG}$ = +30 V (Fig. 8), both dots are "pulled" toward the lower side of the channel and closer to each other. As a result, the current ridges associated with the first added electrons get remarkably tilted, denoting sizeable cross capacitances, and inter-dot capacitive coupling becomes noticeable. In the data of Fig. 8, a sizeable Coulomb interaction between the corner QDs is revealed by the clear separation between (N,M) and (N+1,M+1) charge domains (e.g. between the (0,0) and the (1,1) charge states). The effect on the inter-dot tunnel coupling is harder to appreciate. We investigate that through a numerical calculation carried out for a split-gate device with comparable geometrical parameters (Fig. 9). Figures 10a-c show cross-sectional views of the QD wave functions squared at $V_{BG}$ = -10V, 0V and +5V, respectively. Increasing $V_{BG}$ brings the electron-type QDs toward each other resulting in a stronger tunnel coupling, *t*, as quantitatively shown in Fig. 10d.

In the presence of a small tunnel coupling ($t \sim 1$ μeV), one of the two QDs could be used to encode a qubit and the other one to perform readout, once again leveraging the SB effect. This time, however, readout would not occur through a dc current measurement but through rf gate reflectometry [10,11]. The readout setup is described in Fig. 11. An LC resonator, with resonance frequency $f_{LC}$ in the few-hundred MHz range, is connected to the right gate, confining the readout QD. This connection contributes to a shift in the resonance frequency whose magnitude depends on the quantum capacitance associated with the spin-dependent inter-dot tunneling. Measuring the reflected rf signal gives information on this quantum capacitance and hence on the state of the spin qubit (in the left QD) relative to the reference readout spin (in the right QD), which remains always aligned to the external magnetic field. In principle, this technique can provide fast single-shot readout, and it is potentially scalable.

## IV. PROSPECTS FOR SCALABILITY

Our current knowledge of low-temperature properties of single and double gate MOSFET devices allows us to outline a well-defined development path towards scalable qubit arrays in a linear geometry. The basic idea is presented in the schematic of Fig. 12. In a single nanowire device, an array of closely spaced split-gates defines two parallel lines of corner QDs. One line ($G_{1,x}$ Gates) encodes a linear array of spin qubits with tunneling-mediated nearest-neighbor couplings, the other line ($G_{2,x}$ Gates) consists of spin readout QDs, one for each qubit, where readout is performed via rf gate reflectometry. The targeted gate spacing is 40 nm or lower. Figure 13 shows a preliminary test device with two pairs of face-to-face gates. FDSOI technology offers the additional possibility of gate control from beneath the buried oxide. In particular it would be desirable and perhaps necessary to implement locally tunable tunnel couplings between adjacent QDs. This could in principle be achieved with the aid of locally defined back-gates, with the benefits of a limited cross-capacitance with the Dot-defining front-gates.

The linear geometry of Fig. 12 would provide a useful starting ground for the development of two-dimensional qubit architectures, a necessary requirement for surface-code quantum computation.

ACKNOWLEDGMENT

This work is partly funded by the French Public Authorities through NANO 2017 and Equipex FDSOI1, internal ZPOVA project. The authors also acknowledge financial support from the EU under Project MOS-QUITO (No.688539).

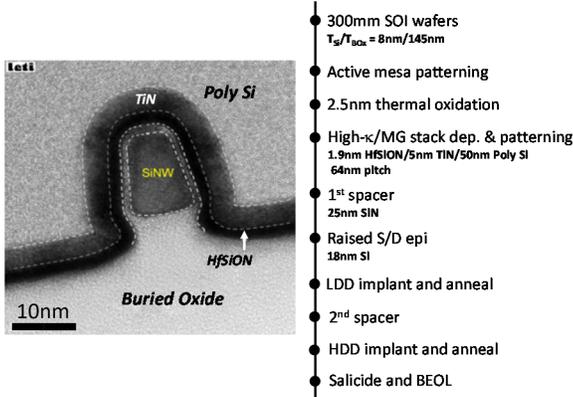

Fig. 1. Simplified process flow and TEM cross-section along the Gate of an SOI NanoWire FET. Wide spacers are primarily used for proper Gate-defined QD confinement, since a thin and undoped SOI region separates it from the carrier reservoirs (see Fig. 2.).

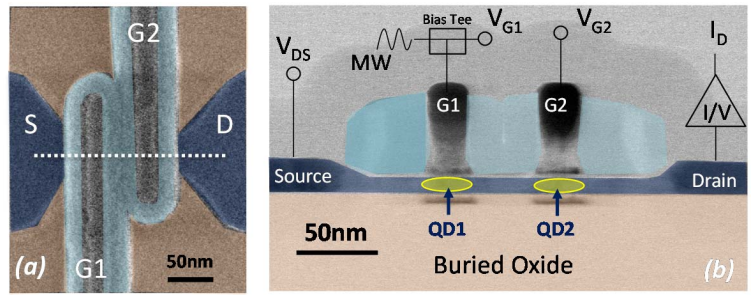

Fig. 2. a) Representative top view of double quantum dot device obtained by scanning electron microscopy (SEM) before the back-end process. The undoped Si channel is 10 nm thick & 20 nm wide. The two gates are 30 nm wide with a 30 nm gap in between. b) Cross-sectional TEM view along the device channel. The wiring scheme for qubit measurements is also shown.

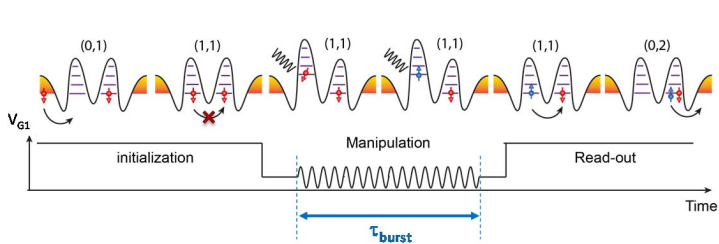

Fig. 3. a) Event sequence in a gate modulation period. Initialization: transport is blocked as soon as a (1,1) triplet state (represented by parallel spins) is loaded. Manipulation: a microwave excitation on Gate 1 induces a coherent rotation of the spin in QD1. Readout: Tunneling out of QD1 into QD2 and then to the Drain contact occurs with a probability proportional to the spin-down component. As a result, Rabi oscillations of the QD1 spin can be detected through oscillations in the measured dc current (see Fig. 4.).

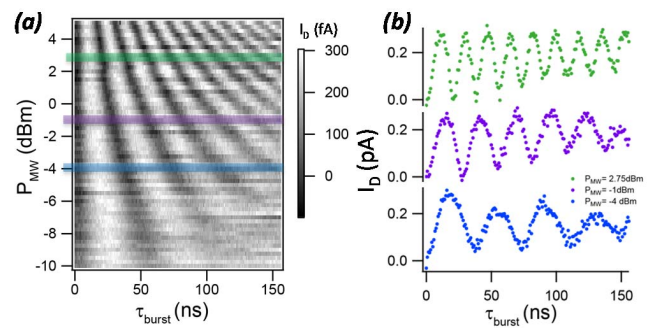

Fig. 4. a) Grey-scale plot of the device current, $I_{sd}$, as a function of MW burst duration, $t_{burst}$, and MW power, $P_{MW}$. Hole-spin coherent rotations are revealed by periodic oscillations of the current as a function of $t_{burst}$. The Rabi rotation frequency increases linearly with $P_{MW}^{1/2}$. b) selected $I_{sd}(t_{burst})$ traces from a).

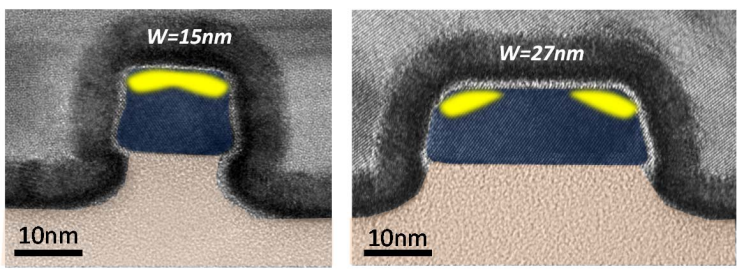

Fig. 5. Cross-sectional TEM views along the Gates of two SOI NWFETs with W=15nm (left) and W=27nm (right). QDs are represented in yellow for qualitative understanding purposes. As the channel width is increased, two separate QDs form in the top corners of the NanoWire.

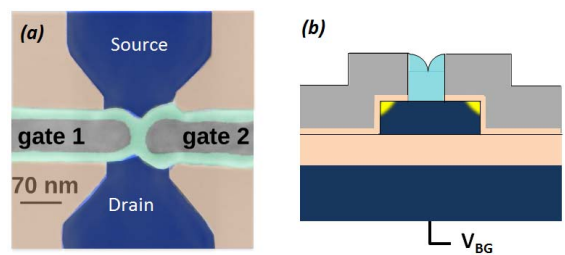

Fig. 6. a) Scanning electron micrograph of a nanowire device with face-to-face top gates. b) Each top gate confines a corner QD, which is addressed individually. A back-gate voltage applied to the substrate provides additional tunability of the interaction between the two corner QDs.

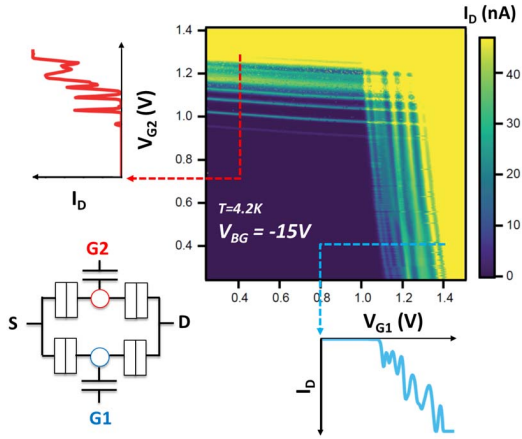

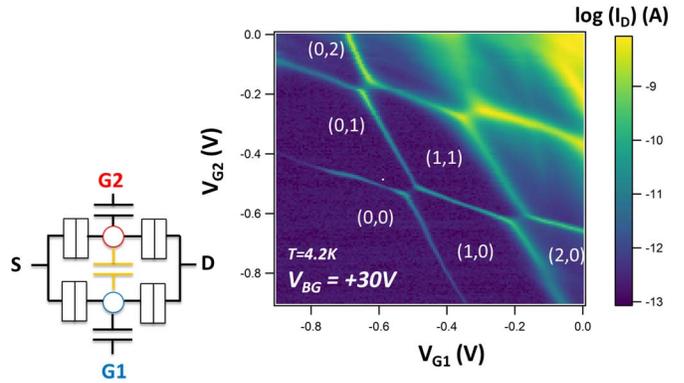

Fig. 7. Linear scale Drain current mapping measured in an n-type split-gate device at 4.2K against $V_{G1}$ and $V_{G2}$, and for a negative $V_{BG}$= -15V. Each QD exhibits Coulomb peaks (cf. $I_D$-$V_G$ cutlines) which are almost independent on the control voltage for the other QD. The dots are well separated, hence the mapping is a superposition of QD1's vertical and QD2's horizontal current ridges.

Fig. 8. Mapping of log($I_D$) in the same device as Fig. 7. ($L_G$=70nm, W=45nm, $S_{GG}$=45nm, $T_{Si}$=11nm, $N_{ch}$=2.10$^{18}$at.cm$^{-3}$, $T_{Box}$=145nm) measured at 4.2K against $V_{G1}$ and $V_{G2}$, for a positive back-Gate bias $V_{BG}$= +30V. The current ridges are tilted, forming a honeycomb pattern delimiting the charge domains. This is a signature of capacitively coupled QDs.

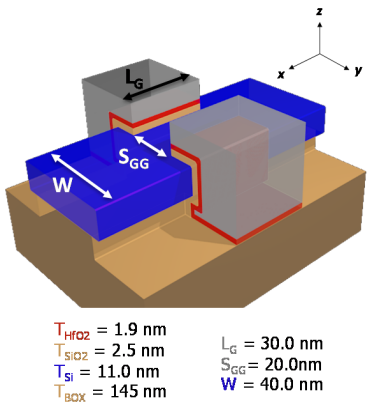

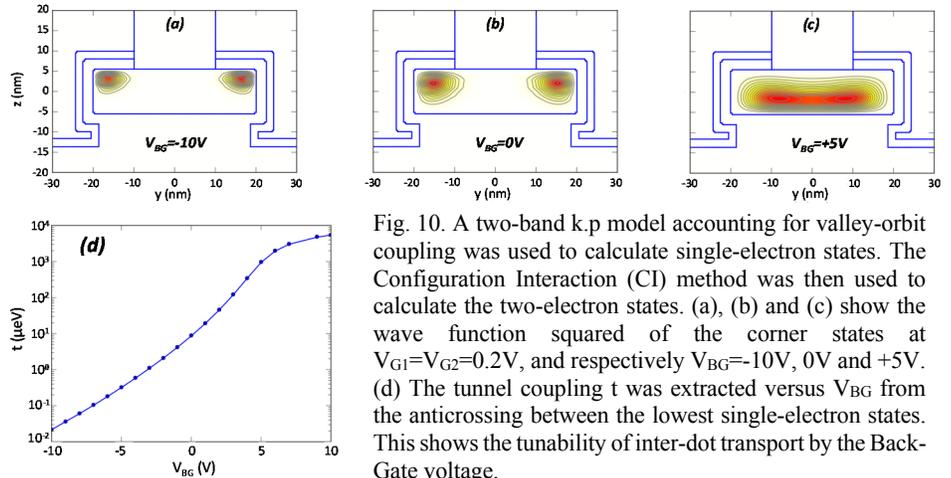

Fig. 9. Structure and dimensions of the n-type face-to-face device with undoped channel simulated in Fig. 10.

Fig. 10. A two-band k.p model accounting for valley-orbit coupling was used to calculate single-electron states. The Configuration Interaction (CI) method was then used to calculate the two-electron states. (a), (b) and (c) show the wave function squared of the corner states at $V_{G1}$=$V_{G2}$=0.2V, and respectively $V_{BG}$=-10V, 0V and +5V. (d) The tunnel coupling t was extracted versus $V_{BG}$ from the anticrossing between the lowest single-electron states. This shows the tunability of inter-dot transport by the Back-Gate voltage.

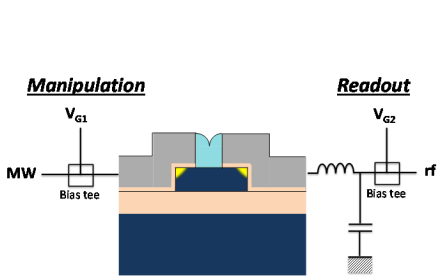

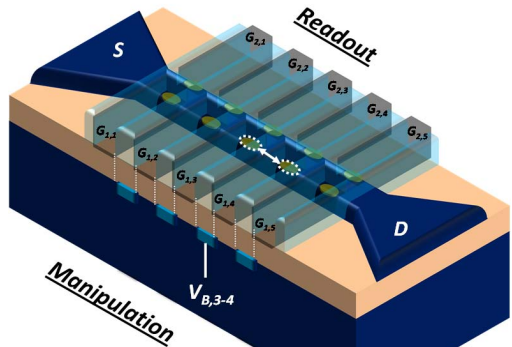

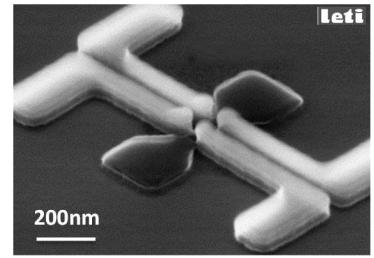

Fig. 11. Principle of the rf reflectometry measurement setup applied to a face-to-face device, enabling scalable, fast single-shot readout of the spin qubit state under G1.

Fig. 12. One-Dimensional array of qubits along a Si NanoWire, using split-gate devices in series. Local Back-Gates formed in the Inter-dots spacings may provide selected tunability of nearest neighbor coupling.

Fig. 13. Tilted top view SEM of two pairs of split-gates on an SOI NanoWire after Gate patterning.